\documentclass{aa}
\usepackage[applemac]{inputenc}
\usepackage{natbib}
\usepackage{txfonts}
\usepackage{graphicx}
\usepackage{color}
\bibpunct{(}{)}{;}{a}{}{,}
\begin{document}

\title{HD depletion in starless cores}
\author{O. Sipilä\inst{1},
	     P. Caselli\inst{2},
	     \and{J. Harju\inst{3,1}}
}
\institute{Department of Physics, PO Box 64, 00014 University of Helsinki, Finland\\
e-mail: \texttt{olli.sipila@helsinki.fi}
\and{School of Physics and Astronomy, University of Leeds, Leeds LS2 9JT, UK}
\and{Finnish Centre for Astronomy with ESO (FINCA), University of Turku, Väisäläntie 20, 21500, Piikkiö,
Finland}
}

\date{Received / Accepted}

\abstract
{}
{We aim to investigate the abundances of light deuterium-bearing species such as HD, $\rm H_2D^+$ and $\rm D_2H^+$ in a gas-grain chemical model including an extensive description of deuterium and spin state chemistry, in physical conditions appropriate to the very centers of starless cores.}
{We combine a gas-grain chemical model with radiative transfer calculations to simulate density and temperature structure in starless cores. The chemical model includes new reaction sets for both gas phase and grain surface chemistry, including deuterated forms of species with up to 4 atoms and the spin states of the light species $\rm H_2$, $\rm H_2^+$ and $\rm H_3^+$ and their deuterated forms.}
{We find that in the dense and cold environments attributed to the centers of starless cores, HD eventually depletes from the gas phase because deuterium is efficiently incorporated to grain-surface HDO, resulting in inefficient HD production on grains for advanced core ages. HD depletion has consequences not only on the abundances of e.g. $\rm H_2D^+$ and $\rm D_2H^+$, whose production depends on the abundance of HD, but also on the spin state abundance ratios of the various light species, when compared with the complete depletion model where heavy elements do not influence the chemistry.}
{While the eventual HD depletion leads to the disappearance of light deuterium-bearing species from the gas phase in a relatively short timescale at high density, we find that at late stages of core evolution the abundances of $\rm H_2D^+$ and $\rm D_2H^+$ increase toward the core edge and the disributions become extended. The HD depletion timescale increases if less oxygen is initially present in the gas phase, owing to chemical interaction between the gas and the dust predecing the starless core phase. Our results are greatly affected if $\rm H_2$ is allowed to tunnel on grain surfaces, and therefore more experimental data not only on tunneling but also on the $\rm O + H_2$ surface reaction in particular is needed.}

\keywords{ISM: abundances -- ISM: clouds -- ISM: molecules -- astrochemistry -- radiative transfer}

\maketitle

\section{Introduction}

Starless cores, sites for potential low-mass star formation, are condensations of dense and cold gas. Owing to the high density and low temperature attributed to these objects, it is expected that species containing heavy elements eventually deplete onto grain surfaces because of their relatively high binding energies. Indeed, depletion of several chemical species such as CO \citep[e.g.][]{Willacy98, Caselli99} and CS \citep[e.g.][]{Tafalla02, Pon09} has been observed toward these objects.

For a long time (since the early 1970s), gas-phase chemical models were the main means of studies of chemical evolution in starless cores. However, even though it has been known for a long time that $\rm H_2$ must be formed on grain surfaces \citep{Gould63, Hollenbach71} and consequently that surface chemistry should play an important role in the chemical evolution of these objects, it is only fairly recently (during the last 10 years or so) that models including the interaction between gas phase and grain surface chemistry have really taken over as the main means of numerical studies of the chemistry in starless cores \citep{Aikawa03, Aikawa05, Garrod07}.

One of the main goals of simulating chemistry in starless cores is not only to explain the observed properties (such as line emission radiation from various species) of these cores, but also to search for new tracer species for the varying physical conditions. Particularly in the very centers of starless cores where density is high, species containing heavy elements are expected to be almost totally depleted onto grain surfaces, and it has been suggested that the light deuterium-bearing species $\rm H_2D^+$ and $\rm D_2H^+$ could serve as the main tracers of these conditions \citep{WFP04, FPW04}, after observations of the ground state rotational transition of ortho-$\rm H_2D^+$ revealed an unexpected strong line toward the prestellar core L1544 \citep{Caselli03}. However, in the so-called complete depletion model of \citet{WFP04} and \citet{FPW04}, in which no heavy elements are present in the gas phase, a description of grain surface chemistry is not included and it has remained somewhat unclear whether the abundances of $\rm H_2D^+$ and $\rm D_2H^+$ would be affected if one were to consider surface chemistry as well.

In this paper, we address this issue by studying the chemistry in physical conditions appropriate to the centers of starless cores using a gas-grain chemical model. To this end, we have constructed new chemical reaction sets for both gas phase and grain surface chemistry that include extensive descriptions of deuterium and spin state chemistry. Our goal is not only to study how the abundances of light deuterated species are affected when grain surface chemistry is taken into account, but also to study spin state abundance ratios and compare these against the predictions of the complete depletion model.

The paper is structured as follows. In Sect.\,\ref{sect2}, we introduce the new chemical reaction sets and the assumptions made in constructing them. We also discuss our chemical and physical models in general. Section \ref{s:results} presents the results of our model calculations. In Sect.\,\ref{s:discussion} we discuss our results and in Sect.\,\ref{s:conclusions} we present our conclusions.

\section{Model}\label{sect2}

We have constructed new chemical reaction sets for both gas phase and grain surface chemistry. The new gas phase reaction set is based on the publicly available Ohio State University {\tt osu\_03\_2008}\footnote{See {\tt  http://www.physics.ohio-state.edu/$\sim$eric/}} (hereafter OSU) reaction set, which was expanded to include the spin states (i.e. ortho and para forms) of the light hydrogen-bearing species $\rm H_2^+$, $\rm H_2$ and $\rm H_3^+$ (here, this process is referred to as ortho/para separation). In addition, species with up to 4 atoms were deuterated using a reaction cloning process. A new surface reaction set was also produced by performing a similar ortho/para separation and deuteration of the surface reaction set of \citet{Sipila12}, itself based on the surface chemistry network of \citealt{Semenov10}; the network includes parameters (e.g. branching ratios, activation energies if applicable) for both ionization (by either external or cosmic ray-induced photons) processes and reactions via the Langmuir-Hinshelwood mechanism (see Sect.\,\ref{ss:chemcode}).

We present below the physical and chemical models used in this study and then discuss the ortho/para separation and deuteration process in detail.

\subsection{Physical model}

\begin{figure}
\centering
\includegraphics[width=\columnwidth]{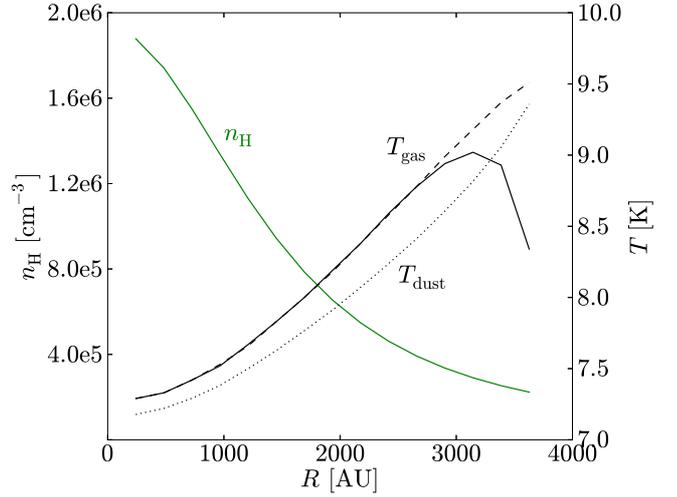}
\caption{The density (green solid line) and temperature (black lines) profiles of the model core. The solid and dashed black lines correspond to the gas temperature at $t = 10^5$ and $t = 10^6$ years, respectively \citep[see also][]{Sipila12}. The dotted line represents the dust temperature.}
\label{fig1}
\end{figure}

In this paper, we present radial abundance profiles of various chemical species. These profiles are produced identically to the method discussed in detail in \citet{Sipila12}: we choose a modified Bonnor-Ebert sphere \citep[e.g.][]{Keto05, Sipila11, Sipila12} as the physical core model and combine chemical and radiative transfer calculations to produce self-consistently calculated radial profiles for both chemical abundances and temperatures ($T_{\rm dust} \neq T_{\rm gas}$ generally). We assume the same cooling species as in \citet{Sipila12}, i.e. $\rm ^{12}CO$ (and its isotopes $\rm ^{13}CO$ and $\rm C^{18}O$), C, O and $\rm O_2$. In this paper we concentrate on comparing our new chemistry model with the complete depletion model, and for this purpose we choose the ``model A'' core of \citet{Sipila12}, that is a modified Bonnor-Ebert sphere of mass 0.25\,$M_{\odot}$. The density and temperature profiles of the model core are plotted in Fig.\,\ref{fig1}. The core has a high average density ($n_{\rm H} \sim 8\times10^5$\,cm$^{-3}$) and a low average temperature ($T \sim 8$\,K, depending slightly on core age), so this core should be appropriate for comparing our new results with the complete depletion model. The gas temperature at the core edge is different for the two core ages because at $t = 10^5$ years CO, the main coolant, is not yet fully depleted at the core edge \citep[see][]{Sipila12}. We note that the gas temperature at the core edge is slightly higher at $t = 10^6$ years than in Sipil\"a (\citeyear{Sipila12}; Fig.\,2). This is due to a slightly lower CO abundance at the core edge brought about by the inclusion of spin state and deuterium chemistry in the present model which modifies slightly the chemistry of undeuterated species compared to the model of \citet{Sipila12}.

We assume that the model core is heavily shielded against external UV radiation -- we set $A_{\rm V} = 10$ mag at the core edge. The cosmic ray ionization rate is set to $\zeta = 1.3\times10^{-17}$\,s$^{-1}$ and we assume spherical grains with radius $a_{\rm g} = 0.1$\,$\mu$m. The assumed initial element abundances with respect to total atomic hydrogen density are given in Table \ref{tab1}. The initial $\rm H_2$ ortho/para ratio is (arbitrarily) set to $1.0 \times 10^{-3}$ (we return to this issue in Sect.\,\ref{ss:opratio}); the initial HD abundance is taken from Sipil\"a et al. (\citeyear{Sipila10}; hereafter S10) and the rest of the initial abundances are taken from \citet{Semenov10}.

\begin{table}
\caption{Initial elemental abundances (see text) with respect to total atomic hydrogen density $n_{\rm H}$.}
\centering
\begin{tabular}{c c}
\hline \hline 
Species & $n(X) / n_{\rm H}$  \\ \hline
$\rm H_2(p)$ & $5.00\times10^{-1}$  \\
$\rm H_2(o)$ & $5.00\times10^{-4}$ \\
$\rm HD$ & $1.60\times10^{-5}$ \\
$\rm He$ & $9.00\times10^{-2}$  \\
$\rm C^+$ & $1.20\times10^{-4}$  \\
$\rm N$ & $7.60\times10^{-5}$ \\
$\rm O$ & $2.56\times10^{-4}$ \\ 
$\rm S^+$ & $8.00\times10^{-8}$ \\
$\rm Si^+$ & $8.00\times10^{-9}$ \\
$\rm Na^+$ & $2.00\times10^{-9}$ \\
$\rm Mg^+$ & $7.00\times10^{-9}$ \\
$\rm Fe^+$ & $3.00\times10^{-9}$ \\
$\rm P^+$ & $2.00\times10^{-10}$ \\
$\rm Cl^+$ & $1.00\times10^{-9}$ \\ \hline
\end{tabular}
\label{tab1}
\end{table} 

\subsection{Chemical model} \label{ss:chemcode}

The chemical code used in this study is essentially the same as the one discussed in \citet{Sipila12}, with some minor bug fixes. Gas phase and grain surface reactions are solved simultaneously using the rate equation method, and gas-grain interaction occurs through adsorption and thermal desorption processes. The rate coefficient for adsorption is $k_i^{\rm ads} = {\upsilon}_i \, S \sigma$, where ${\upsilon}_i = \sqrt{8 k_{\rm B} T_{\rm gas} / \pi m_i}$ is the thermal speed of species $i$, $S$ is the sticking coefficient (set to unity for all species) and $\sigma = \pi a_{\rm g}^2$ is the grain cross section. Desorption occurs mainly\footnote{Thermal desorption as a separate process is also included in the model, but this is negligible in the temperatures considered here.} through transient heating of the grains by cosmic rays, with rate coefficient $k_{\rm CR} = f \, k_{\rm TD} (70 \, {\rm K})$, where $k_{\rm TD}$ is the thermal desorption rate coefficient \citep{HH93}. Photodesorption is not included in the present model; we return briefly to this issue in Sect.\,\ref{ss:hdoabu}.

Species adsorbed onto grain surfaces are assumed to be physisorbed and react via the Langmuir-Hinshelwood mechanism. Following the formalism of \citet{HHL92}, the rate coefficient for surface reactions is given by
\begin{equation}
k_{ij} = \alpha \, \kappa_{ij} \left( R_i^{\rm diff} + R_j^{\rm diff} \right) / n_d \,
\end{equation}
where $\alpha$ is the branching ratio in the case that the reaction has multiple product channels, $\kappa_{ij}$ is the reaction probability, $R_i^{\rm diff}$ is the diffusion rate of species $i$ on the grain surface and $n_d$ is the density of grains. In addition to being destroyed in two-body reactions, species on the surface can be photodissociated either by external UV photons ($k^{\rm phot} = \alpha \exp(-\gamma / A_{\rm V})$) or by cosmic ray induced UV photons ($k^{\rm CRphot} = \alpha\zeta$), although the former process is negligible in the high extinction environment assumed here.

The explicit expressions of $\kappa_{ij}$ and $R_i^{\rm diff}$ depend on whether or not quantum tunneling is assumed to occur. In case of thermal diffusion
\begin{equation}
R_i^{\rm diff} = {\nu_i \over N_s} \exp(-E_i^{\rm diff} / T_{\rm dust}) \, ,
\end{equation}
where $\nu_i = \sqrt{2 n_s k_{\rm B} E_i^{\rm b} / \pi^2 m_i}$; $N_s = n_s \, 4\pi a_{\rm g}^2$ is the number of binding sites on the grain and $E_i^{\rm b}$ and $E_i^{\rm diff}$ are the binding and diffusion energies, respectively. As in \citet{Sipila12}, we assume $n_s = 1.5 \times 10^{15}$\,cm$^{-2}$ for the surface density of binding sites and $E_i^{\rm diff} = 0.77 \, E_i^{\rm b}$ for the relation between diffusion and binding energy. If a reaction is exothermic and has no activation energy, $\kappa_{ij} = 1$, but for exothermic reactions with activation energy, $\kappa_{ij} = \exp(-E_{\rm a} / T_{\rm dust})$ where $E_{\rm a}$ is the activation energy of the reaction. If quantum tunneling of light species (H and D) is assumed (see also Sect.\,\ref{ss:tunneling}), the reaction probability and diffusion rate are -- in the reactions involving these species -- replaced by
\begin{equation}
\kappa_{ij} = \exp \left[ -2 \, ( a / \hbar) \, (2 \, \mu \, k_{\rm B} E_a)^{1/2} \right]
\end{equation}
and
\begin{equation}
R_i^{\rm diff,q} = {\nu_i \over N_s} \exp \left[-2 \, ( a / \hbar) \, (2 \, m \, k_{\rm B} E_i^{\rm diff})^{1/2} \right] \, ,
\end{equation}
respectively \citep{HHL92}, where $\mu$ is the reduced mass of the reactants. We assume $a = 1\,\AA$ for the barrier width. We note that the above assumption of a rectangular barrier is not valid for endothermic reactions. Some of the reactions with activation barriers included in the surface reaction set may be endothermic. However, because reactions with high barriers turn out not to influence our results significantly, we have chosen to omit this possible problem with the tunneling probability.

Contrary to \citet{Sipila12}, we assume 500\,K for the binding energy of $\rm H_2$, which is in the range of typical values assumed by other authors, usually somewhere between 430\,K \citep{Garrod11} and 600\,K \citep{Cazaux10}. In \citet{Sipila12}, the binding energy of $\rm H_2$ was arbitrarily decreased to 100\,K to avoid the problem of producing unphysical amounts of surface $\rm H_2$ \citep[this issue has been discussed by e.g.][]{Garrod11}. As we now consider a binding energy value of 500\,K, the surface $\rm H_2$ abundance is typically high in our models; however, due to the low temperature (see below) considered here, a high surface $\rm H_2$ abundance should not present a problem as the surface reactions involving $\rm H_2$ have high activation energies (and $\rm H_2$ is not allowed to tunnel).

As in \citet{Sipila12}, we assume a binding energy of 1390\,K for atomic oxygen \citep{Bergeron08, Cazaux11}. Apart from $\rm H_2$ and O, the binding energies for undeuterated species are adopted from \citet{Garrod06}. Following the usual approach in the literature \citep[e.g.][]{Cazaux10, Taquet13}, we assume that for each deuterated species, the binding energy is equal to that of the corresponding undeuterated species. We discuss the binding energies further in Sect.\,\ref{ss:binding}.

\subsection{Ortho/para separation}

To add the spin states of $\rm H_2^+$, $\rm H_2$ and $\rm H_3^+$ into the OSU reaction set, we first extracted the reactions involving these species from the OSU database. A Python script was written to analyze each reaction and to add the spin states using a predetermined set of rules.

Most reactions are separated according to spin selection rules. As an example, consider the reaction
\begin{equation}\label{reac1}
{\rm H_3^+} + {\rm O} \mathop{\longrightarrow}\limits^k {\rm OH^+} + {\rm H_2},
\end{equation}
with rate coefficient $k$. This reaction has three possible pathways:
\begin{eqnarray}\label{h3+sep}
{\rm H_3^+(o)} &+& {\rm O} \mathop{\longrightarrow}\limits^k {\rm OH^+} + {\rm H_2(o)} \nonumber \\
{\rm H_3^+(p)} &+& {\rm O} \mathop{\longrightarrow}\limits^{{1\over2}k} {\rm OH^+} + {\rm H_2(o)} \\
{\rm H_3^+(p)} &+& {\rm O} \mathop{\longrightarrow}\limits^{{1\over2}k} {\rm OH^+} + {\rm H_2(p)} \nonumber \, ;
\end{eqnarray}
the branching ratios can be calculated with e.g. the method of Oka (\citeyear{Oka04}; see also Appendix \ref{appendixa}). We note that in the literature, alternative approaches to the separation exist -- for example, \citet{Flower06a} assumed branching ratios of 1:2 for the separation of $\rm H_3^+(p)$ in reactions analogous to (\ref{h3+sep}), while \citet{Pagani09} assumed 1:1, as is done here.

The above branching ratios are not applied to all reaction types. One example of this is charge transfer reactions where we have assumed that the ortho/para forms are conserved in the reaction. Also, whenever $\rm H_2$ is formed in a reaction containing only reactants other than $\rm H_2^+$, $\rm H_2$ and $\rm H_3^+$, such as in the reaction
\begin{equation}\label{spec_example}
\rm NH^+ + H_2O \longrightarrow HNO^+ + H_2 \, ,
\end{equation}
we assume that $\rm H_2$ is formed in its para state (see Appendix \ref{appendixa}). This is a simplifying assumption, ensuring that we do not have to follow the spin states of every species with multiple protons. There are several special cases besides reaction (\ref{spec_example}) -- they are listed in Appendix \ref{appendixa} along with our assumptions.

There are approximately 1000 reactions containing $\rm H_2^+$, $\rm H_2$ or $\rm H_3^+$ in the OSU reaction file; performing the ortho/para separation results in $\sim$1700 reactions, so that the amount of new reactions brought about by the separation process is not significant (with respect to the total size of the reaction network, which is about 11600 reactions after ortho/para separation and deuteration; see Sect.\,\ref{sect2.2}). A similar separation process was applied to the surface reaction set; in this case the separation process adds only about 50 new reactions.

\begin{figure*}
\centering
\includegraphics[width=17cm]{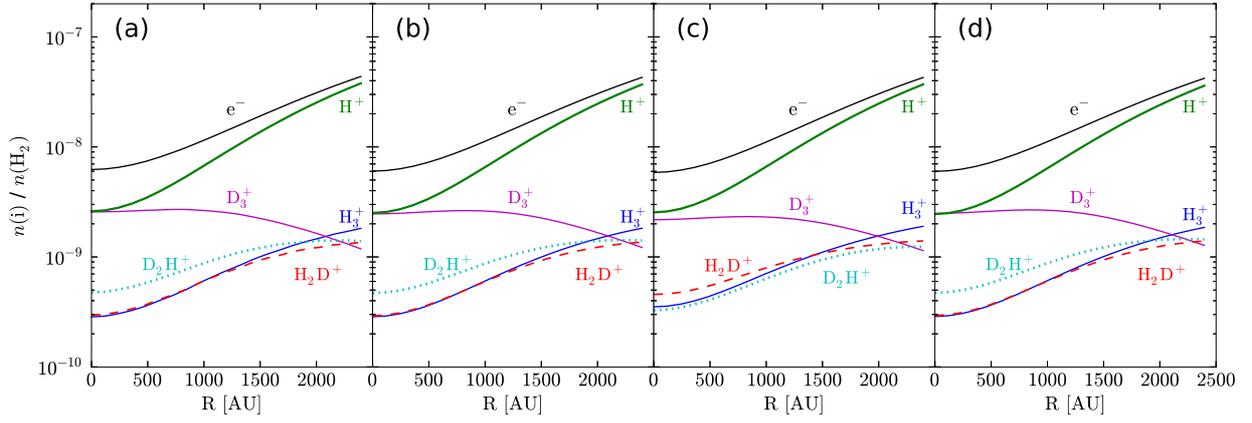}
\caption{Radial abundances (with respect to $n(\rm H_2)$) of protons, $\rm H_3^+$ and its deuterated isotopologues and electrons (indicated in the figure). Panel {\bf (a)} is a reproduction of the middle panel of Fig.\,6 in \citet{Sipila10} (showing abundances at $t = 10^6$ years), calculated with the version of the chemical code used in that paper (see text). The two middle panels use the same physical model as in panel {\bf (a)}, but chemical evolution has been calculated with the present version of the chemical code (see text) with quantum tunneling turned either on (panel {\bf b}) or off (panel {\bf c}). Panel {\bf (d)} plots again the same physical situation but calculated with the new reaction network presented in this paper, with initial heavy element abundances set to zero.}
\label{fig2}
\end{figure*}

\subsection{Deuteration process}\label{sect2.2}

After the ortho/para separation process was carried out, we produced deuterated versions of both the gas phase and surface reaction sets. In practice, deuterons are substituted in place of protons in each reaction and combinatorial arguments are invoked (as in the case of ortho/para separation) to work out branching ratios, where applicable. For example, deuterating the reaction
\begin{equation}
{\rm NH^+} + {\rm H_2O} \mathop{\longrightarrow}\limits^k {\rm NH_2^+} + {\rm OH}
\end{equation}
results in 9 new reactions including, for example, the reactions
\begin{eqnarray}
{\rm NH^+} &+& {\rm HDO} \mathop{\longrightarrow}\limits^{{2\over3}k}  {\rm NHD^+} + {\rm OH} \nonumber \\
{\rm NH^+} &+& {\rm HDO} \mathop{\longrightarrow}\limits^{{1\over3}k}  {\rm NH_2^+} + {\rm OD} \, ,
\end{eqnarray}
where complete scrambling, i.e. a statistical approach, is assumed. Our deuteration routine is similar to that presented in \citet{Rodgers96} -- recently an analogous method has also been applied to the {\tt osu\_2009} network by \citet{Albertsson11}. As pointed out by \citet{Rodgers96}, the complete scrambling assumption is not appropriate for all reactions, particularly for those involving more complex molecules with OH endgroups, for example. However, as we are interested in the chemistry of relatively simple species in the present paper, we have chosen a rather relaxed approach toward this problem and assume that the complete scrambling assumption is generally valid. Of course, there are a number of special cases; for example, in charge transfer reactions no atoms are interchanged, and the deuteration routine is designed to make sure this does not occur in the deuterated version of the reaction either. Other important special cases are deuterated analogues of reactions such as
\begin{equation}
\rm H_3^+ + H_2O \longrightarrow H_3O^+ + H_2 \ .
\end{equation}
We have assumed that reactions such as this proceed by $\rm H_3^+$ donating a proton and thus, for example, the reaction $\rm D_2H^+$ + $\rm H_2O$ cannot result in $\rm HD_2O^+$~+~$\rm H_2$.

Deuterated reactions involving only the lightest species (up to helium) are taken from previous works. The deuteration chemistry of the light species in the context of a complete depletion model (no heavy elements in the gas phase, and no surface chemistry except for the formation of $\rm H_2$, HD and $\rm D_2$) was discussed originally by \citet{WFP04} and \citet{FPW04}, and subsequently by many authors. The goal of the present paper is to study how the inclusion of heavy elements and a description of gas-grain interaction including surface chemistry affects the gas-phase chemistry of $\rm H_3^+$ and its deuterated isotopologues; we have therefore incorporated the reaction set used in S10 into the expanded OSU set so that the final reaction set is consistent with the complete depletion model as far as the chemistry of light species is concerned. In practice, we replaced the appropriate reactions in the new reaction set with their S10 counterparts and added any reactions from S10 that did not exist in the new model\footnote{For example, the OSU set does not include the reaction $\rm H_3^+$ + $\rm H_2$ $\rightarrow$ $\rm H_3^+$ + $\rm H_2$, but when considering spin states (and deuterated forms) explicitly, the variants of this reaction need to be added to the model. This important system has been recently analyzed by \citet{Hugo09}.}. This allows for a maximally consistent comparison between the two cases.

We have similarly added to the final (deuterated and spin state-separated) surface reaction set the reactions and the associated activation energies included in the models of \citet{Cazaux10} and \citet{Cazaux11}.

As a final step, all reactions in the final deuterated gas phase and surface reactions sets were checked for elemental balance, i.e. that both sides of all reactions contain the same amount of elements. The final gas phase reaction set, including deuterium and spin state chemistry, contains about 11600 reactions. The final surface reaction set contains about 1350 reactions.

\section{Results}\label{s:results}

\subsection{Comparison of the new chemical code against the S10 code}\label{ss:comparison}

A previous (gas-phase chemistry only) version of our chemical code was used in S10 to study deuterium chemistry in the complete depletion limit; to test the new reaction set discussed here, and the chemical code itself, we reproduced some of the results of S10 using both the old and the new versions of the chemical code (Fig.\,\ref{fig2}). Panel {\bf (a)} is a reproduction of the middle panel in Fig.\,6 in S10 (abundances are shown at $t = 10^6$ years), calculated using the old version of the chemical code and adopting the same density and temperature profiles, the same reaction set and assuming the same values for the physical parameters as in S10. It should be noted that because only gas-phase chemistry was considered in S10, the grain-surface formation mechanisms for $\rm H_2$, HD and $\rm D_2$ were hard-coded into the program and no surface species as such were included. The new version of the code, however, includes surface chemistry as described above and in \citet{Sipila12}; to compare these two approaches, we plot in panels {\bf (b)} and {\bf (c)} the results of calculations otherwise identical to those of panel {\bf (a)}, but calculated using the new code with quantum tunneling either included (panel~{\bf b}) or excluded (panel~{\bf c}). It can be seen that without quantum tunneling, the results of S10 are not reproduced. The reason for this is the very low temperature ($\sim$6.5\,K in the center) of the model core; without quantum tunneling, surface production of HD is inefficient at this temperature and eventually leads to a drop in gas phase HD abundance. This in turn increases the $\rm H_3^+$ and $\rm H_2D^+$ abundances and decreases the $\rm D_2H^+$ abundance, while the $\rm D_3^+$ abundance is relatively unaltered. The reasons for these effects are discussed in, e.g., \citet{FPW04}, \citet{Flower06a} and \citet{Pagani09}. With quantum tunneling included, the new code gives virtually identical results compared to the old code because of the rapid conversion of surface H and D to HD. It should be noted that the creation rate of HD on grain surfaces depends critically on the assumed temperature \citep[e.g.][]{Cazaux09}; already at $\sim$8\,K, the difference between the tunneling and no tunneling cases is very small as demonstrated by comparing results at radii $\gtrsim 2000$\,AU in panels ({\bf b}) and ({\bf c}) of Fig.\,\ref{fig2}, where the temperature is $\sim$ 7.5\,K.

\begin{figure*}
\centering
\includegraphics[width=17cm]{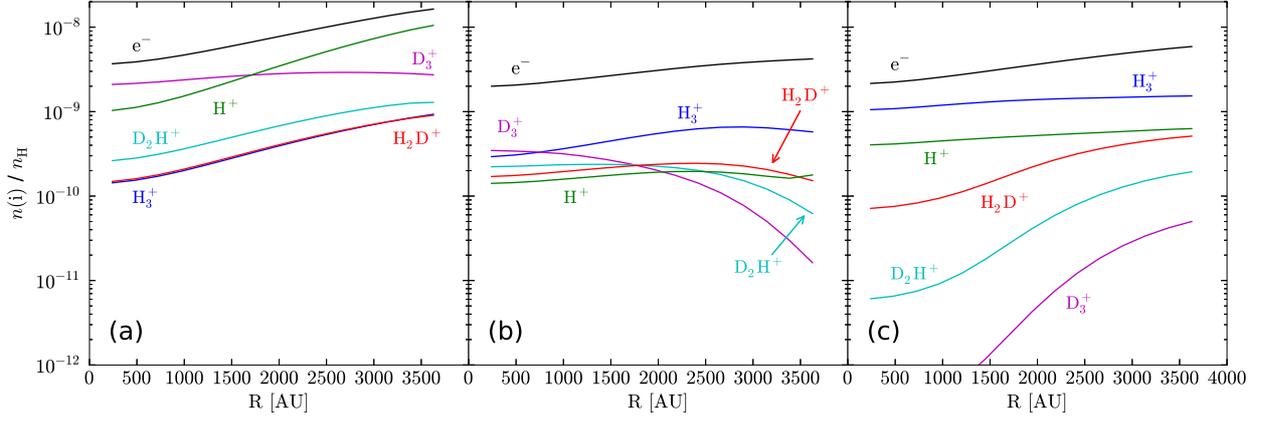}
\caption{Radial abundances (with respect to $n_{\rm H}$) of protons, electrons and $\rm H_3^+$ and its deuterated isotopologs (indicated in the figure). Panel ({\bf a}) corresponds to the complete depletion model at $t = 10^6$\,years (using the reaction set of S10), panels ({\bf b}) and ({\bf c}) correspond to the present model at $t = 10^5$\,years and $t = 10^6$\,years, respectively.}
\label{fig3}
\end{figure*}

In panel {\bf (d)} we plot again the same physical situation as in panels {\bf (a-c)}, using the new reaction set described above but with initial heavy element abundances set to zero (and quantum tunneling included). As would be expected since the light element chemistry originates mainly from the S10 set (see Sect.\,\ref{sect2.2}), the abundances are practically identical to those presented in panels {\bf (a)} and {\bf (b)}. Even though the deuteration process described above introduces some reactions involving only the light species that do not exist in the S10 set, it turns out that these reactions (such as some reactions involving the $\rm H^-$ and $\rm D^-$ anions) are rather insignificant (at least for this set of physical conditions).

In what follows, all modeling results correspond to the new ortho/para separated and deuterated reaction set with quantum tunneling switched on, unless otherwise stated.

\subsection{Comparison with the complete depletion model} \label{ss:cdlimit}

In the complete depletion model, it is assumed that ``heavy'' species (i.e. those containing elements heavier than He) are frozen onto grain surfaces at high gas densities, allowing the deuteration chemistry of $\rm H_3^+$ to proceed unhindered. The deuteration chain of $\rm H_3^+$ is highly dependent on the abundance of HD \citep[e.g.][]{WFP04}, which is produced on grain surfaces and is thus dependent on the available amount of atomic H and D. In the complete depletion model, grain-surface atomic H and D are spent only in the reactions creating $\rm H_2$, $\rm HD$ or $\rm D_2$; in essence, it is assumed that heavy species on grain surfaces are locked in unreactive species, such as water or methanol. However, in a model where heavy elements are not totally depleted, the large amount of reacting species on the surface may reduce the amount of HD produced because of the additional reaction pathways for H and D (compared to the complete depletion model), and thus affect the abundances of $\rm H_3^+$ and its deuterated isotopologues in the gas phase.

\begin{figure*}
\centering
\includegraphics[width=17cm]{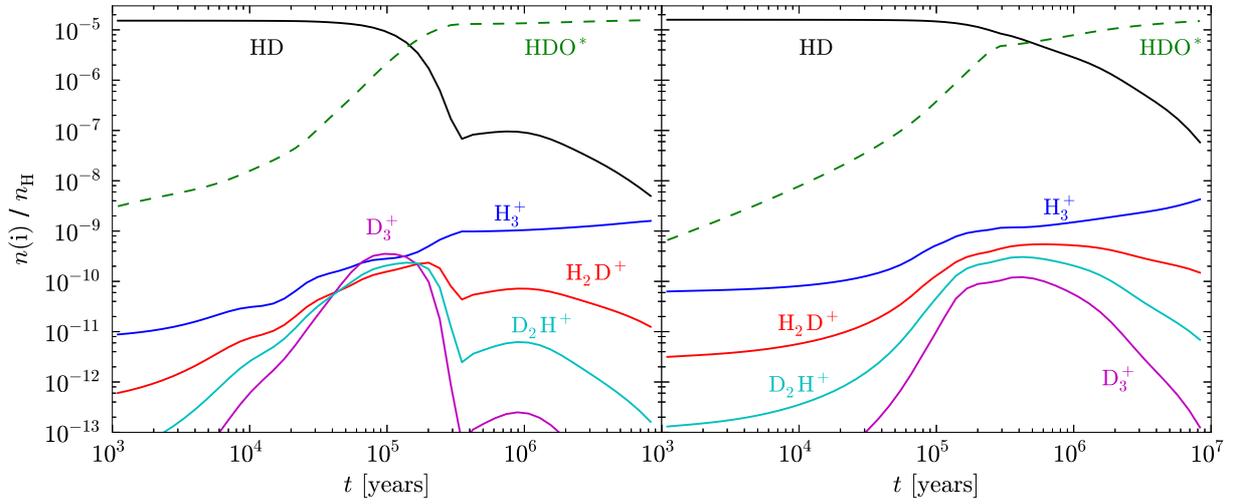}
\caption{Abundances (with respect to $n_{\rm H}$) of selected species (indicated in the figure) as functions of time in the innermost shell ($R \sim 240$\,AU, $n_{\rm H} \sim 1.9\times10^6$\,cm$^{-3}$; left panel) and the outermost shell ($R \sim 3600$\,AU, $n_{\rm H} \sim 2.2\times10^5$\,cm$^{-3}$; right panel) of the model core.}
\label{fig4}
\end{figure*}

To demonstrate this, we plot in Fig.\,\ref{fig3} the radial abundances of $\rm H_3^+$ and its deuterated isotopologs using the reaction set of either S10 (panel {\bf a}) or this work (panels {\bf b} and {\bf c}); also plotted are the protons and electrons. Panels ({\bf a}) and ({\bf c}) correspond to a core age of $10^6$ years and panel ({\bf b}) corresponds to $t = 10^5$ years. It can be immediately seen that the abundances change dramatically depending on the adopted reaction set and that in the present model the abundances are highly time-dependent even at advanced core ages (in the complete depletion model, the abundances reach steady-state across the core at about $t = 2\times10^5$ years). In the dense center of the core, where one would expect a high degree of deuteration of $\rm H_3^+$ based on the complete depletion model \citep[e.g.][]{FPW04, Pagani09}, the present model predicts a completely opposite situation at $t = 10^6$ years -- in this case, the degree of deuteration increases toward the edge of the core, i.e. toward lower density. This can be understood through the surface chemistry; we plot in Fig.\,\ref{fig4} the abundances of HD, HDO$^{\ast}$ (in this paper, an asterisk represents a surface species) and $\rm H_3^+$ and its deuterated isotopologs as a function of time in the innermost (left panel) and outermost (right panel) shells of the model core. In the dense center of the core ($n_{\rm H} \sim 10^6$\,cm$^{-3}$), H$^{\ast}$ and D$^{\ast}$ react preferentially with O$^{\ast}$ to produce OH$^{\ast}$ and OD$^{\ast}$, which react again with H$^{\ast}$ and D$^{\ast}$ to produce $\rm H_2O^{\ast}$ and HDO$^{\ast}$ (these processes are efficient because oxygen is initially in atomic form; see Table~\ref{tab1} and Sect.\,\ref{ss:iniabu}). This process almost completely suppresses the formation of HD$^{\ast}$ through the reaction of H$^{\ast}$ with D$^{\ast}$ (see also Sect.\,\ref{ss:tunneling}). Thus, since its abundance is not efficiently regenerated on grains, HD ultimately depletes from the gas phase. HD depletion coincides with heavy element depletion -- this is because the $\rm H_3^+$ abundance increases with heavy element depletion and starts to destroy HD, which is not efficiently created on grain surfaces owing to the reasons mentioned above.

The HD abundance experiences a slight increase at about $t = 3.5 \times 10^5$ years. This arises because OH$^{\ast}$ and OD$^{\ast}$ disappear from the grain surfaces as they are processed into $\rm H_2O^{\ast}$ and HDO$^{\ast}$; this frees up some H$^{\ast}$ to react with HDS$^{\ast}$ and HDCO$^{\ast}$ to produce HS$^{\ast}$ + HD$^{\ast}$ and HCO$^{\ast}$ + HD$^{\ast}$, respectively (see also Sect.\,\ref{ss:tunneling}). The HD$^{\ast}$ thus formed desorbs and momentarily regenerates the HD abundance.

At the lower density of the core edge ($\sim 10^5$\,cm$^{-3}$), represented by the right panel of Fig.\,\ref{fig4}, HD depletes  less than at very high density ($\sim 10^6$\,cm$^{-3}$), implying a higher deuteration degree of $\rm H_3^+$ at an advanced core age. The slower depletion of HD at the lower density is due to the overall depletion timescale being longer in these conditions, which suppresses deuterium chemistry (as seen in the behaviour of $\rm H_3^+$ deuteration at early times), and thus there is somewhat less D$^{\ast}$ available to form OD$^{\ast}$ (and $\rm HDO^{\ast}$ through OH$^{\ast}$ + D$^{\ast}$). As a consequence, the timescale for deuterium incorporation into HDO$^{\ast}$ is longer than at high density. Also in this case, total HD depletion is prevented by the reactions H$^{\ast}$~+~HDS$^{\ast}$ and H$^{\ast}$~+~HDCO$^{\ast}$.

\begin{figure*}
\centering
\includegraphics[width=17cm]{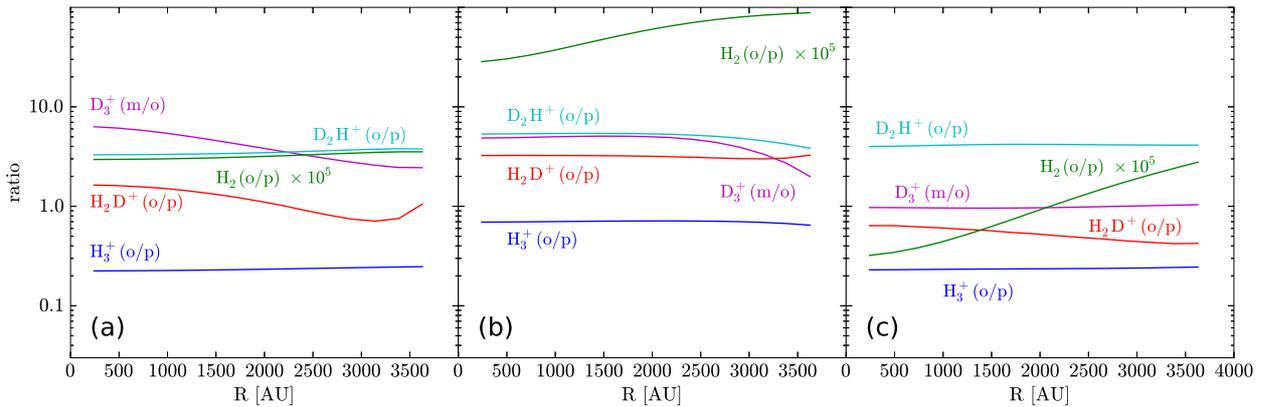}
\caption{Nuclear spin state abundance ratios of $\rm H_3^+$ and its deuterated isotopologs and of $\rm H_2$ (multiplied by $10^5$). Panel ({\bf a}) corresponds to the complete depletion model at $t = 10^6$\,years, panel ({\bf b}) to the present model at $t = 10^5$\,years and panel ({\bf c}) to the present model at $t = 10^6$\,years.}
\label{fig5}
\end{figure*}

Figures \ref{fig3} and \ref{fig4} demonstrate that the deuteration degree of $\rm H_3^+$ is highly dependent on time. Indeed, at $t = 10^5$ years, the agreement of the present model with the complete depletion model is much better than at $t = 10^6$ years. With the exception of $\rm D_3^+$, for which we obtain much lower abundances than in the complete depletion model, the agreement of the two models is rather good in the core center at $t = 10^5$ years, but less so at the core edge where depletion is not as efficient. The detailed behavior of the deuteration degree depends on many factors, including the binding energies of various species, but the main result of Figs. \ref{fig3}~and~\ref{fig4} is the depletion of HD which does not occur in the complete depletion model.

From an observational standpoint it is interesting that the radial abundances of $\rm H_2D^+$ and $\rm D_2H^+$ are similar to each other in the new model at $t = 10^5$ years (see Fig.\,\ref{fig3}), although their abundances are clearly lower in the new model than in the complete depletion model at the core edge. This agrees with observational results that the two species are present with comparable abundances \citep{Parise11, Vastel12}.

It is also observed in Fig.\,\ref{fig3} that the ionization degree decreases when one adds heavy elements to the model. This is because in the present model, electrons are mainly removed in dissociative recombination reactions with $\rm HCO^+$ and $\rm HCNH^+$; the rate coefficients of these reactions are about a factor of 3 higher than those of $\rm H_3^+$ and its deuterated forms, leading to a lower ionization degree in a model with heavy elements when compared to the one with zero heavy element abundances.

\subsection{Ortho/para ratios}

\begin{figure*}
\centering
\includegraphics[width=17cm]{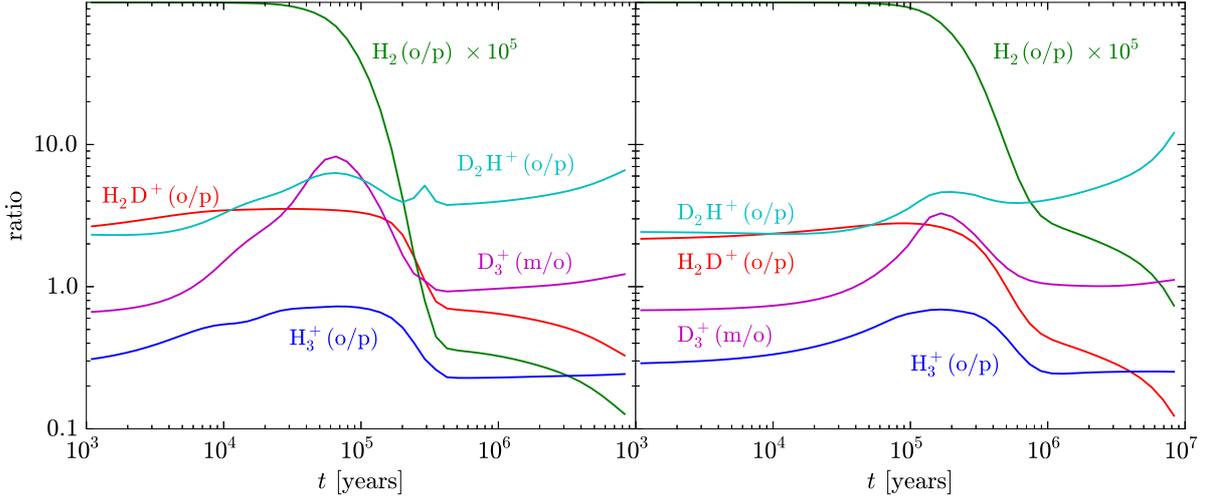}
\caption{Nuclear spin state abundance ratios of $\rm H_3^+$ and its deuterated isotopologs and of $\rm H_2$ (multiplied by $10^5$) as functions of time in the innermost shell ($R \sim 240$\,AU, $n_{\rm H} \sim 1.9\times10^6$\,cm$^{-3}$; left panel) and the outermost shell ($R \sim 3600$\,AU, $n_{\rm H} \sim 2.2\times10^5$\,cm$^{-3}$; right panel) of the model core.}
\label{fig6}
\end{figure*}

In Figs. \ref{fig2}, \ref{fig3} and \ref{fig4}, we have plotted the abundances of $\rm H_3^+$ and its deuterated isotopologs as sums of the abundances of their respective nuclear spin states. To illustrate how the ortho/para ratios of these species differ in the present model compared with the complete depletion model, we plot in Fig.\,\ref{fig5} the ortho/para ratios of $\rm H_3^+$, $\rm H_2D^+$, $\rm D_2H^+$ and $\rm H_2$ (multiplied by $10^5$) and the meta/ortho ratio of $\rm D_3^+$ as predicted by both the complete depletion model and the new model presented here. The panels depict the same core ages as in Fig.\,\ref{fig3}. Figure \ref{fig5} is supplemented by Fig.\,\ref{fig6}, where the spin state ratios are plotted as a function of time in the innermost and outermost shells of the model core. As could be expected based on the evolution of the HD abundance in the present model, the spin state abundance ratios of the various species, which depend on the abundances of HD and ortho-$\rm H_2$ \citep{WFP04, FPW04, Sipila10}, are somewhat different in the two models, depending also on core age. The $\rm H_3^+$, $\rm H_2D^+$ and $\rm D_2H^+$ spin state ratios are similar (within a factor of a few) in both models, particularly at $t = 10^5$ years, but there are larger differences in the behavior of the $\rm D_3^+(m/o)$ ratio and the $\rm H_2(o/p)$ ratio depending on the model.

One can see from Figs. \ref{fig5} and \ref{fig6} that the o/p ratios (and the spin temperatures, $T_{\rm spin}$) of $\rm H_2$, $\rm H_3^+$, and $\rm H_2D^+$ decrease as the core evolves. The effect is particularly marked for $\rm H_2$, with $T_{\rm spin}$ dropping below 15\,K after $10^6$ years in both the innermost and the outermost shells of the model core. Also for $\rm D_2H^+$ and $\rm D_3^+$, a slight decrease in $T_{\rm spin}$ (i.e. an increase in the o/p and m/o ratios, respectively) can be seen at late times (after $10^6$ years) when compared with early times ($<10^4$ years). The $\rm D_3^+(m/o)$ ratio settles at $\sim 1$ which indicates a clearly super-thermal $T_{\rm spin}$ of 42\,K, but the ratio experiences a peak at around $10^5$ years with a m/o ratio of $\sim 8$ ($T_{\rm spin}~\sim~16$\,K). For $\rm H_3^+$, $\rm H_2D^+$ and $\rm D_2H^+$, the spin temperatures settle between 15 and 30\,K at late stages of core evolution.

The $\rm D_3^+$ (m/o) ratio is much lower in the core center at $t = 10^6$ years in the present model than in the complete depletion model. At the high density of the core center, the ratio first begins to increase following heavy element depletion (initiating the deuteration chemistry), but then decreases again due to HD depletion --  comparing Figs.\,\ref{fig4}~and~\ref{fig6} reveals that a decrease of the $\rm D_3^+$ (m/o) ratio indeed coincides with HD depletion. This is because meta-$\rm D_3^+$ is most efficiently created in a spin-state-conversion reaction of ortho-$\rm D_3^+$ with HD and in the reaction between $\rm D_2H^+(o)$ and HD; therefore HD depletion leads to a decrease of the $\rm D_3^+$ (m/o) ratio.

The $\rm H_2$ (o/p) ratio is very different in the present model than in the complete depletion model. The effect of HD depletion can be seen here as well; $\rm H_2(o)$ is most efficiently created in the gas phase in the reactions
\begin{equation}
\rm H_3^+ + HD \longrightarrow H_2D^+ + H_2 \, ,
\end{equation}
\begin{equation}
\rm H_2D^+ + HD \longrightarrow D_2H^+ + H_2 \, ,
\end{equation}
and
\begin{equation}
\rm D_2H^+ + HD \longrightarrow D_3^+ + H_2 \, ,
\end{equation}
so that HD depletion directly affects $\rm H_2(o)$ creation. HD depletion is stronger and occurs earlier at the core center than at the edge, and this accounts for the differences seen between panels ({\bf b}) and ({\bf c}) in Fig.\,\ref{fig5}. Note that the gas phase production of $\rm H_2(o)$ is now the most important factor controlling the $\rm H_2(o/p)$ ratio because quantum tunneling is included; see Sect.\,\ref{ss:tunneling}.

Interestingly, the present model predicts very similar $\rm H_2D^+$ and $\rm D_2H^+$ (o/p) ratios at $t= 10^5$ years compared with the complete depletion model even at the core edge, where the $\rm H_2D^+$ and $\rm D_2H^+$ (ortho+para) abundances differ from the complete depletion model (Fig.\,\ref{fig3}). Finally we note that the ortho/para ratios predicted by our models at $t = 10^6$ years are for the most part compatible with recent modeling results of Vastel et al. (\citeyear{Vastel12}; their complete depletion case), although our chemical network is much larger and the present model includes a time-dependent treatment of depletion.

\section{Discussion}\label{s:discussion}

\subsection{Initial heavy element abundances}\label{ss:iniabu}

In this paper, we assume that the gas is initially atomic, with the exceptions of hydrogen being in molecular form and deuterium being locked in HD (see Table \ref{tab1}). In other words, we do not take into account any possible chemical history of the core, i.e. the possibility that some of the heavier elements could be locked in ices at the start of the calculation. This can be particularly important for HDO$^{\ast}$ which is created from OH$^{\ast}$ and OD$^{\ast}$ and is thus dependent on the amount of available atomic oxygen. The assumed initial HD abundance corresponds roughly to the average gas phase D/H abundance ratio, ${\rm (D/H)_{gas}}$, in the Local Bubble \citep{Wood04}. Substantially lower values of ${\rm (D/H)_{gas}}$ have been determined towards lines of sight with large column densities. \citet{Linsky06} find a correlation between deuterium depletion and the depletion of metals, and suggest that D can be incorporated in carbonaceous dust grains. The assumption about the initial gas phase deuterium abundance determines the overall degree of deuteration, but is not likely to affect the relative abundances of deuterated species. 

\begin{figure}
\centering
\includegraphics[width=\columnwidth]{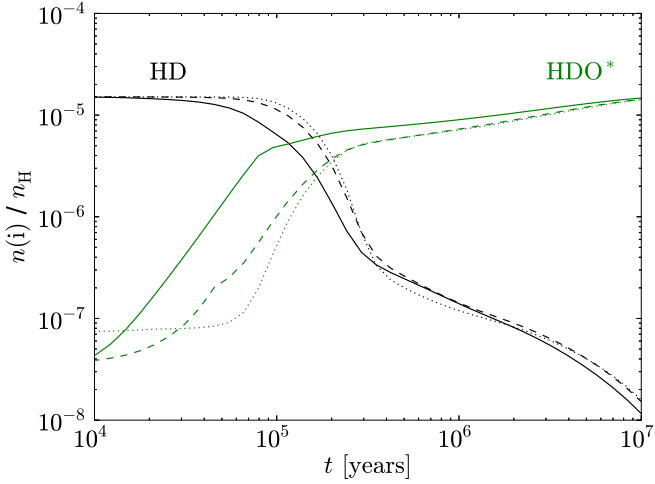}
\caption{The abundances of HD and HDO$^{\ast}$ as a function of time at a density of $n_{\rm H} = 2\times10^6$\,cm$^{-3}$ calculated for progenitor cloud (see text) ages of $10^4$ (solid lines), $10^5$ (dashed lines) or $10^6$ (dotted lines) years.}
\label{fig7}
\end{figure}

Because the present model does not include a description of core collapse (from, e.g., a diffuse cloud into the dense configuration that is the present model core), we have investigated the effect of the initial abundances on HD depletion by first calculating chemical evolution in a low-density cloud with $n_{\rm H} = 2\times10^3$\,cm$^{-3}$, visual extinction $A_{\rm V} = 3$~mag and temperature $T_{\rm gas} = T_{\rm dust} = 10$\,K \citep[which should be appropriate for these values of density and visual extinction,][]{Juvela11}. The abundances of all species corresponding to three different times ($10^4$, $10^5$ and $10^6$ years) were then extracted and used as initial abundances for the dense core. Figure \ref{fig7} shows the result of such calculations in a single-point chemical model with $n_{\rm H} = 2\times10^6$\,cm$^{-3}$, temperature $T_{\rm gas} = T_{\rm dust} = 7.5$\,K and $A_{\rm V} = 100$~mag, corresponding roughly to the interior of the model core discussed in this paper. It is observed that the older the progenitor cloud, the longer it takes for deuterium to get locked into HDO$^{\ast}$. This is because after $10^6$ years of progenitor cloud evolution, about a half of the initial oxygen abundance has been converted to $\rm H_2O^{\ast}$, which limits somewhat the surface reactions leading to the production of $\rm HDO^{\ast}$ in the dense core phase (the $\rm HDO^{\ast}$/$\rm H_2O^{\ast}$ ratio is only about $10^{-3}$ at the end of the progenitor cloud evolution). Notably, very little HD depletion takes place in the diffuse cloud. We also tested the effect of more extreme initial abundances by locking all oxygen initially in $\rm H_2O^{\ast}$ (50\%) and $\rm CO^{\ast}$ (50\%) and all nitrogen in $\rm NH_3^{\ast}$ -- in these calculations, HD still depletes (by about two orders of magnitude in $\sim 10^6$ years) because surface D gets locked into multiple singly deuterated surface species ($\rm NH_2D^{\ast}$, $\rm HDCO^{\ast}$ and $\rm HDO^{\ast}$).

While the details of the chemistry change when one starts from molecular initial abundances (with respect to the model with initially atomic heavy element abundances), our main conclusion (HD depletion) remains unaffected. Figure~\ref{fig7} demonstrates this for $n_{\rm H} = 2\times10^6$\,cm$^{-3}$, but we have also performed test calculations at lower densities to confirm that HD depletion is largely unaffected in those conditions as well.

\subsection{Initial $\rm H_2$ ortho/para ratio}\label{ss:opratio}

\begin{figure}
\centering
\includegraphics[width=\columnwidth]{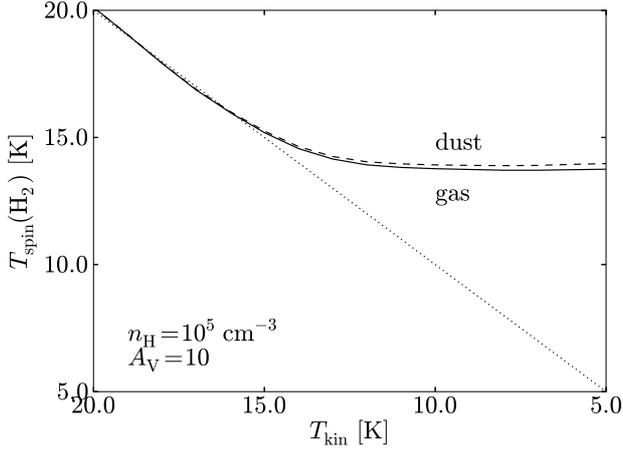}
\caption{The spin temperatures of gas-phase (solid line) and grain-surface (dashed line) $\rm H_2$ as a function of the kinetic temperature for $n(\rm H_2) = 10^5$\,cm$^{-3}$ and $A_{\rm V} = 10$~mag.}
\label{fig8}
\end{figure}

Dense cores condense from diffuse molecular clouds where the observationally derived ortho/para column density ratios of $\rm H_2$ indicate spin temperatures in the range 50--70\,K \citep[$\rm H_2(o/p) \sim 0.3-0.8$,][]{Crabtree11}. \citet{Crabtree11} showed that the $\rm H_2$ spin temperatures in diffuse clouds correspond to the gas kinetic temperatures because the ortho/para ratio of $\rm H_2$ is thermalized through collisions with $\rm H^+$. Also in dense clouds with elevated temperatures, the spin-state conversion reactions
\begin{equation}
{\rm H^+} + {\rm H_2(o)} \rightarrow {\rm H^+} + {\rm H_2(p)}
\end{equation}
and
\begin{equation}\label{reac11}
{\rm H^+} + {\rm H_2(p)} \rightarrow {\rm H^+} + {\rm H_2(o)}
\end{equation}
govern the $\rm H_2(o/p)$ ratio. As discussed by Flower et al. (\citeyear{Flower06a}; their Sect. 3.1), the spin temperature of $\rm H_2$ follows closely the kinetic temperature down to $T_{\rm kin} \sim 20$\,K. The $\rm H_2(o/p)$ ratio corresponding to this temperature is $1.8\times10^{-3}$. Below 20\,K, the endothermic para-ortho conversion reaction with $\rm H^+$ (reaction \ref{reac11}) starts to lose importance, and $\rm H_2(o)$ production is dominated by formation on grains. The consequence is that with a decreasing kinetic temperature, the $\rm H_2$ spin temperature rises above the kinetic temperature. In cold, dense cores ($T_{\rm kin}\sim 10$\,K), the competition between the more favored ortho production on grains and the efficient para-ortho conversion in reactions with $\rm H^+$ and $\rm H_3^+$ (and its deuterated counterparts) in the gas phase leads to an $\rm H_2$ spin temperature which lies a few K above the kinetic temperature. This point is illustrated in Fig.\,\ref{fig8} at high values of gas density ($n(\rm H_2) = 10^5$\,cm$^{-3}$) and visual extinction ($A_{\rm V} = 10$~mag); the gas (solid line) and dust (dashed line) $\rm H_2$ spin temperatures lie clearly above the kinetic temperature (dotted line) at low values of $T_{\rm kin}$.

Based on the ideas presented above we have chosen, somewhat arbitrarily, the initial value  $1.0\times10^{-3}$ for the $\rm H_2(o/p)$ ratio in our chemistry model (see Table \ref{tab1}). This ratio corresponds to an $\rm H_2$ spin temperature of 18.7\,K. Although the inital ortho/para ratio depends on the thermal history of the core, the spin temperature should in any case lie somewhere between the current kinetic temperature and 20\,K, because during core condensation and cooling, $\rm H_2$ has probably been thermalized down to 20\,K. Therefore, we think the initial value quoted above is more realistic than the thermal value at 10\,K ($3.6\times10^{-7}$) or the statistical ratio at high temperatures (3) which is the assumed ortho/para formation ratio on grains. The choice of the initial abundances has some effect during the first $\sim 10^5$ years in our simulations; at late times they are forgotten because of the high density of the model core.

\begin{figure*}
\centering
\includegraphics[width=17cm]{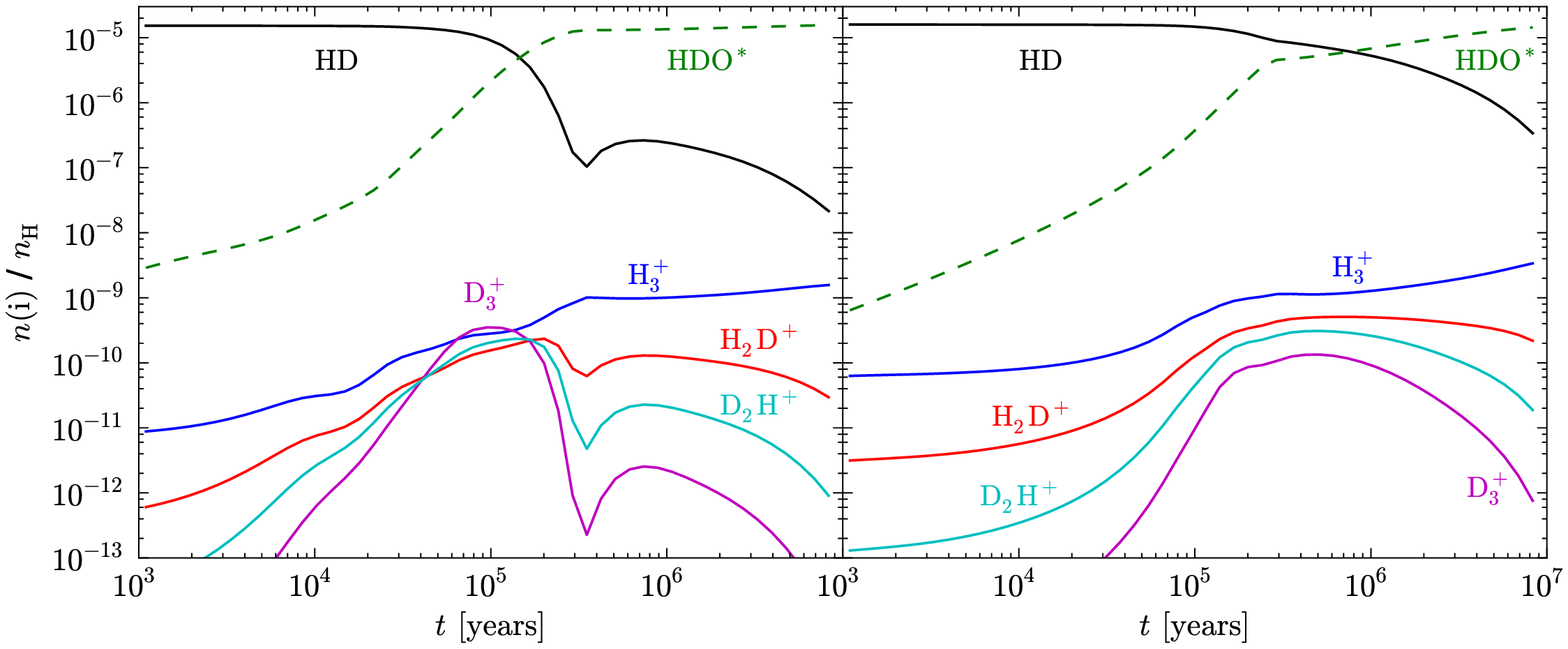}
\caption{As Fig.\,\ref{fig4}, but in the calculations quantum tunneling is not included.}
\label{fig9}
\end{figure*}

\begin{figure*}
\centering
\includegraphics[width=17cm]{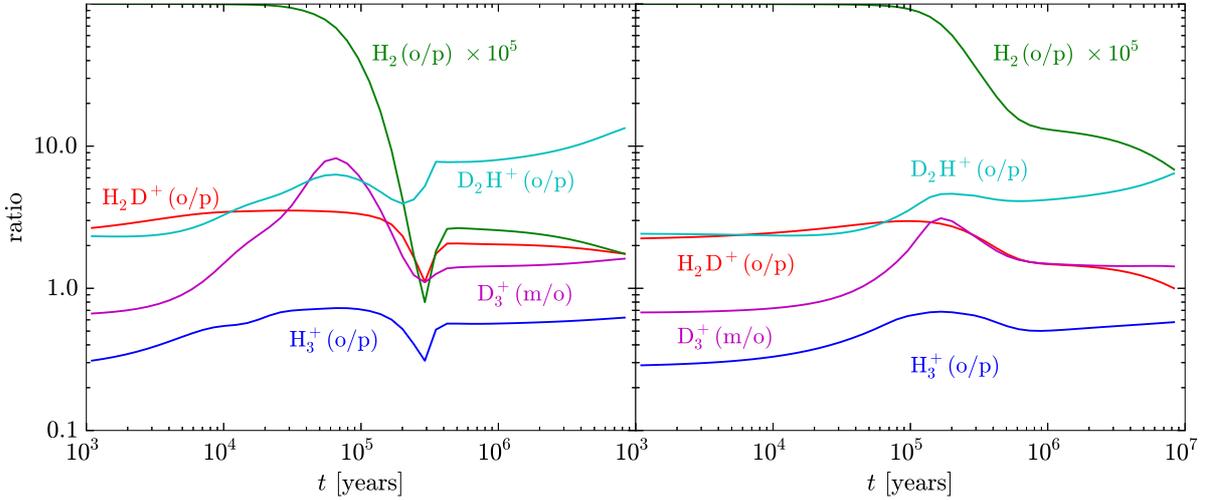}
\caption{As Fig.\,\ref{fig6}, but in the calculations quantum tunneling is not included.}
\label{fig10}
\end{figure*}

\subsection{Quantum tunneling}\label{ss:tunneling}

Whether diffusion of species via quantum tunneling occurs on grain surfaces or not is still a subject of some debate in modern astrochemistry. There is some experimental evidence that tunneling does not occur \citep[e.g.][]{Katz99} and on the other hand some modeling results indicate that tunneling should be taken into account \citep[e.g.][]{Cazaux04}. In this paper, we have included quantum tunneling so that our analysis would be maximally consistent with the complete depletion model (see also Sect.\,\ref{ss:comparison} and Fig.\,\ref{fig2}). However, it is also prudent to take into account the possibility that tunneling does not occur and to see whether HD depletion is affected by this assumption; with this in mind, we have carried out calculations similar to those presented in the earlier sections with quantum tunneling turned off. The results of these calculations are presented in Fig.\,\ref{fig9}, which should be compared with Fig.\,\ref{fig4}. It is observed that while the detailed behavior of the abundances as a function of time changes when quantum tunneling is turned off, the main result of this work -- the ultimate depletion of HD at high density -- is not affected.

The details of the (deuterium) chemistry change when quantum tunneling is turned off, which can be explained as follows. When one assumes quantum tunneling, many of the surface reaction pathways for H and D that would be otherwise unavailable due to high activation energy become accessible because tunneling increases the reaction probability and the effective rate coefficient by increasing the diffusion rate (see Sect.\,\ref{ss:chemcode}). This effectively reduces the abundances of H$^{\ast}$ and D$^{\ast}$ so that it becomes increasingly difficult to form $\rm H_2^{\ast}$ and HD$^{\ast}$ through the basic reactions H$^{\ast}$~+~H$^{\ast}$ and H$^{\ast}$~+~D$^{\ast}$, respectively. Indeed, the main pathway for producing HD$^{\ast}$ at advanced core ages is H$^{\ast}$~+~HDCO$^{\ast}$ (as discussed in Sect.\,\ref{ss:cdlimit}) -- when tunneling is included. However, without tunneling, all of the channels for which $E_a \neq 0$ are effectively closed off owing to the very low temperature, and  HD$^{\ast}$ is mainly produced in H$^{\ast}$~+~D$^{\ast}$ after OH$^{\ast}$ and OD$^{\ast}$ are processed into $\rm H_2O^{\ast}$ and HDO$^{\ast}$\footnote{Since the $\rm H^{\ast} + O^{\ast}$ and $\rm D^{\ast} + O^{\ast}$ reactions are barrierless, the inclusion of tunneling does not change the reaction probability, but does increase the effective rate coefficient. However, these reactions are efficient both with and without tunneling owing to the large $\rm O^{\ast}$ abundance.}. This reaction is more efficient than H$^{\ast}$~+~HDCO$^{\ast}$ in the tunneling case and thus gives a slightly higher HD$^{\ast}$ yield, and for this reason HD$^{\ast}$ depletes slightly slower in the model with no tunneling (also observed in the form of a somewhat more pronounced bump in the HD abundance at high density); the decreased HD depletion also slightly increases the abundances of the deuterated isotopologs of $\rm H_3^+$ at late times.

The model without tunneling also yields a larger amount of $\rm H_2(o)$ at late times compared to the model with tunneling, which is demonstrated in Fig.\,\ref{fig10} (to be compared with Fig.\,\ref{fig6}). The reason for this is again the increased H$^{\ast}$ abundance which leads to efficient $\rm H_2(o)$ creation through the H$^{\ast}$~+~H$^{\ast}$ reaction. When tunneling is included, $\rm H_2(o)$ is not efficiently created on grain surfaces because 1) the H$^{\ast}$~+~H$^{\ast}$ reaction is suppressed, and 2) other surface reactions preferentially create $\rm H_2(p)$ owing to the adopted ortho/para separation rules (see Appendix~\ref{appendixa}). An observationally important consequence is the increased $\rm H_2D^+(o/p)$ ratio at late times when tunneling is turned off.

To summarize, the inclusion or exclusion of quantum tunneling modifies the detailed behavior of the deuterium chemistry and particularly the spin state ratios, but the main result of this paper (eventual HD depletion at high density) is not affected. While our assumptions regarding the production of $\rm H_2(o)$ naturally influence its abundance in the model, more (experimental) evidence is needed to justify either including or excluding quantum tunneling on grain surfaces so that the detailed behavior of the spin state abundances at advanced core ages can be better constrained.

\subsection{Binding energies}\label{ss:binding}

Experimental evidence suggests that the binding energies of light species on grain surfaces depend strongly on the properties of the ice \citep[e.g.][]{Hornekaer05, Amiaud06}. Because in a realistic situation the properties of the surface (porosity, content of the ice) change as a function of time, one should consider time-dependent binding energies for the various species instead of the constant values considered here (and in the majority of other works). Without an appropriate model, it is difficult to quantify whether HD depletion would be affected if this were done, but to study the influence of varying binding energies we have run some test calculations using a single-point chemical model corresponding to the conditions in the center of the model core, varying the binding energies of H and $\rm H_2$ and their deuterated forms between 300\,K and 600\,K \citep[which should cover the typical values on either non-porous or porous amorphous solid water;][]{Taquet13}. In all test calculations, HD eventually depletes from the gas phase, however the depletion is somewhat less pronounced with increasing H and D binding energy when tunneling is not included (because a higher H or D binding energy translates to lower surface reaction rates so that HDO$^{\ast}$ is not formed as efficiently). If tunneling is included, the variation in binding energy does not influence HD depletion in the range of binding energies studied (unless $\rm H_2$ is allowed to tunnel, see Sect.\,\ref{ss:hdoabu}).

In this work, we assume that the binding energy of each deuterated species is the same as that of the undeuterated analogue. This is an approximation since in principle, one would expect deuterated species to be somewhat more strongly bound to grain surfaces than their undeuterated counterparts due to their slightly higher mass \citep{Tielens83, Perets05, Kristensen11}. However, we do not expect taking the binding energy difference between non-deuterated and deuterated species into account to influence the main result (HD depletion) of this paper; if we assumed higher binding energies for deuterated species, this would probably only lead to more efficient HD depletion owing to decreased desorption efficiency. 

\subsection{The $\rm HDO^{\ast}$ abundance}\label{ss:hdoabu}

\begin{figure*}
\centering
\includegraphics[width=17cm]{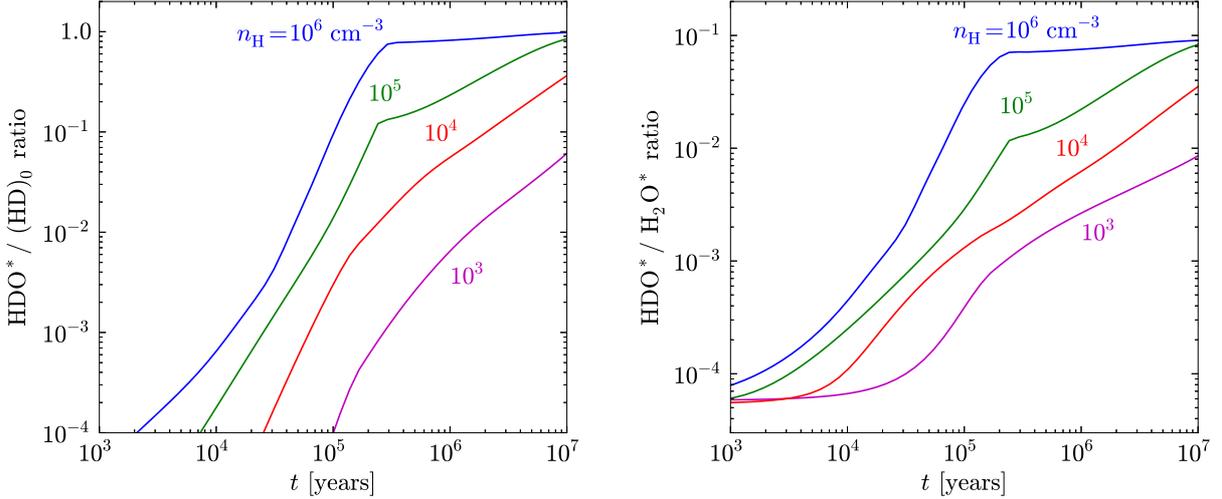}
\caption{{\sl Left:} The ratio of the HDO$^{\ast}$ abundance to the {\sl initial} gas phase HD abundance at different densities (indicated in the figure) as a function of time. {\sl Right:} The $\rm HDO^{\ast} / H_2O^{\ast}$ ratio at different densities as a function of time.}
\label{fig11}
\end{figure*}

Looking at Fig.\,\ref{fig4}, it seems that the conversion of HD into $\rm HDO^{\ast}$ depends only very loosely on density. In these high-density cases, oxygen is finally locked mainly in $\rm H_2O^{\ast}$ while deuterium is locked in $\rm HDO^{\ast}$. However, at lower densities, the recycling of deuterium is slower and the late-time abundance of $\rm HDO^{\ast}$ is consequently lower than at high density. This is illustrated in the left panel of Fig.\,\ref{fig11}, where the ratio of $\rm HDO^{\ast}$ to the initial gas phase HD abundance at different densities is plotted as a function of time. For these calculations, we have assumed $T_{\rm gas} = T_{\rm dust} = 10$\,K and $A_{\rm V} = 10$~mag. Evidently, the relative amount of deuterium locked in $\rm HDO^{\ast}$ at low density is small even at advanced core ages. Because $\rm HDO^{\ast}$ is the most abundant D-bearing species on grain surfaces in our model, it follows that HD is hardly depleted at all at low density.

In the right panel of Fig.\,\ref{fig11}, we plot the $\rm HDO^{\ast} / H_2O^{\ast}$ ratio at different densities as a function of time. This plot can be compared with other studies of water ice deuteration, such as the recent study by \citet{Taquet13}; although our model is in many respects different from theirs, we predict similar values for the $\rm HDO^{\ast} / H_2O^{\ast}$ ratio for typical core lifetimes \citep[$10^5 - 10^6$ years; see Figs. 8 and 10 in][]{Taquet13}. The agreement is very good considering the differences between our model and theirs; for instance, \citet{Taquet13} consider fixed values of the $\rm H_2$ ortho/para ratio, whereas in our model the ratio is calculated as a function of time. Furthermore, in our model the binding energies of the various species do not vary with the $\rm H_2$ content on the surface (see Sect.\,\ref{ss:chemcode}), which should make some difference on the progression of the surface chemistry, although this is difficult to quantify without a direct comparison of the two models (see also Sect.\,\ref{ss:binding}). Finally, our model (unlike that of \citealt{Taquet13}) does not include a treatment of photodesorption. Consequently, the desorption efficiency is somewhat lower in our model which may translate, in particular, to slightly higher $\rm HDO^{\ast}$ abundances in the present model than in the model of \citet{Taquet13}. It should be noted that it is unclear if HD depletion occurs in a multilayer model, such as that of \citet{Taquet13}. Based on our results, HD depletion occurs because deuterium is mostly locked in $\rm HDO^{\ast}$, i.e. the late-time $\rm HDO^{\ast}$ abundance is similar to the initial HD abundance. In the model of \citet{Taquet13}, the $\rm H_2O^{\ast}$ and $\rm HDO^{\ast}$ abundances are about an order of magnitude lower than in our model in similar physical conditions, so that the entire deuterium reservoir in HD cannot be transferred solely to $\rm HDO^{\ast}$. However, in a multilayer model deuterium might still be efficiently locked into multiple surface species (e.g. ${\rm H_2O}$, ${\rm H_2CO}$, ${\rm NH_3}$, ...) which could influence the gas-phase HD abundance -- this issue should be investigated.

\citet{Aikawa12} have recently studied deuterium chemistry in the context of a collapse model, and they predict $\rm HDO^{\ast} / H_2O^{\ast}$ ratios of $\sim 0.01$. Their model does not include spin state chemistry, and thus gives an upper limit on the strength of deuterium fractionation (which is reduced by $\rm H_2(o)$; \citealt{Flower06a}; \citealt{Taquet13}). In the model of \citet{Aikawa12}, the sticking coefficient is set to 0.5 whereas we use a value of unity, which means that the relative amount of reactive species on the surface should be larger in our model at any given time. Finally, in the model of \citet{Aikawa12}, the model core spends a relatively short time in the prestellar phase (the protostar is born in $2.5\times10^5$ years) -- comparing the $\rm HDO^{\ast} / H_2O^{\ast}$ ratio given by their model near the core edge where $n_{\rm H} \sim 10^5$\,cm$^{-3}$ and $T \sim 10$\,K \citep[Figs.~1~and~3 in][]{Aikawa12} with our results (Fig.\,\ref{fig11}), the agreement is again rather good.

\begin{figure*}
\centering
\includegraphics[width=17cm]{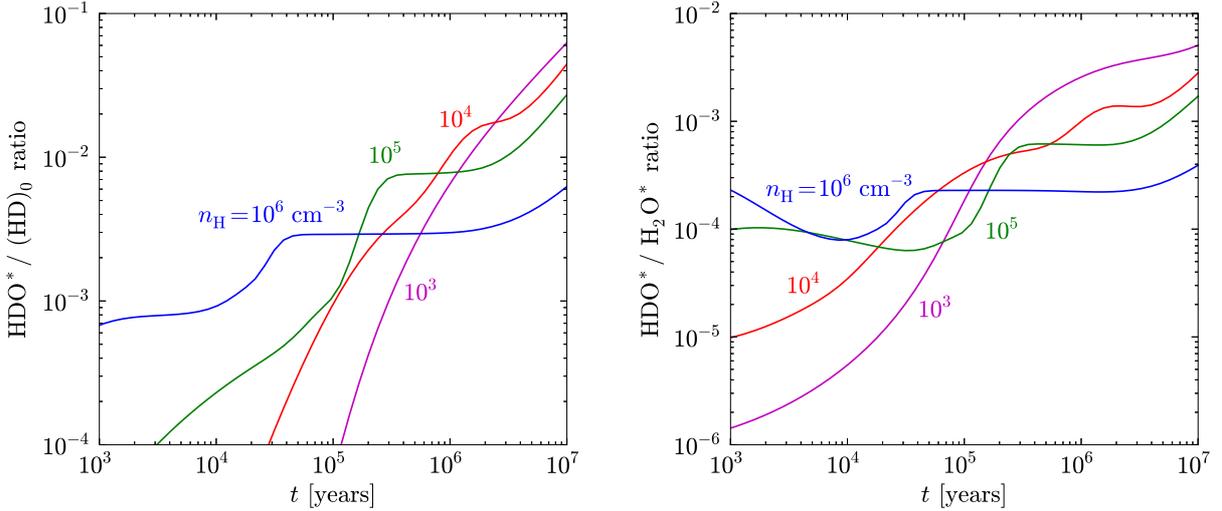}
\caption{As Fig.\,\ref{fig11}, but assuming $\rm H_2$ tunneling (in addition to H and D).}
\label{fig12}
\end{figure*}

\citet{Cazaux11} reported modeling results yielding very low $\rm HDO^{\ast} / H_2O^{\ast}$ ratios at low temperatures (and a strong temperature dependence of the ratio). In the model of \citet{Cazaux11}, $\rm H_2$ is allowed to tunnel and this allows $\rm H_2O^{\ast}$ to be efficiently produced in the $\rm O^{\ast} + H_2^{\ast}$ reaction despite its high activation barrier \citep[3000\,K in][]{Cazaux11}, which in their model leads to a low $\rm HDO^{\ast} / H_2O^{\ast}$ ratio at low temperatures. We have performed test calculations including the reactions of \citet{Cazaux11} in the chemical model and allowing $\rm H_2$ to tunnel as well (in addition to H and D) -- the results of these calculations are plotted in Fig.\,\ref{fig12}, which assumes the same physical parameters as in Fig.\,\ref{fig11}. Evidently, $\rm HDO^{\ast}$ is produced less efficiently at increasing densities. This is because $\rm H_2$ tunneling allows the barriers associated with the reactions involving $\rm H_2$ to be more easily overcome, and the $\rm O^{\ast} + H_2^{\ast}$ reaction (in particular) becomes important. Consequently, $\rm O^{\ast}$ and $\rm OH^{\ast}$ are used up fast and the $\rm HDO^{\ast} / H_2O^{\ast}$ ratio levels off before the gas-phase deuterium chemistry, releasing atomic D, takes over. Thus, we find low $\rm HDO^{\ast} / H_2O^{\ast}$ ratios ($< 10^{-3}$) even at high density, which leads to little HD depletion. We note that while this is an interesting result, more experimental data not only on tunneling on grain surfaces but also on the $\rm O^{\ast} + H_2^{\ast}$ reaction at low temperatures is needed to investigate this issue. \citet{Oba12} claim to find no evidence for this reaction to proceed at low temperature. However, because of the difficulty to carry out laboratory experiments with atomic oxygen, more experiments are needed to confirm the Oba et al. result. Finally, the results of Fig.\,\ref{fig12} are also affected by the $\rm H_2$ binding energy; for example for $E_{\rm H_2}^{\rm b} = 300$\,K, surface $\rm H_2$ desorbs fast which greatly decreases the rates of the reactions involving $\rm H_2$, and we get results closer to Fig.\,\ref{fig11}. Therefore, a model including time-dependent binding energies, possibly yielding less $\rm H_2^{\ast}$ than the static-binding-energy model considered here, could also yield high $\rm HDO^{\ast}$ abundances at high density even when $\rm H_2$ tunneling is included. This point warrants further investigation.

\subsection{Observational constraints on the $\rm HDO^{\ast}$ abundance}

The depletion of heavy elements and the accompanying increase in the atomic D/H abundance ratio in the gas phase is expected to result in high degrees of deuteration in water, formaldehyde, and methanol incorporated in the icy mantles of grains \citep{Tielens83}. Observations suggest, however, that deuteration is clearly more pronounced in ${\rm H_2CO^{\ast}}$ and ${\rm CH_3OH^{\ast}}$ than in ${\rm H_2O^{\ast}}$ (see e.g. \citealt{Taquet13}, and the references therein). This has been explained by the fact that water ice is formed at an early stage when CO depletion is not marked. Observations of the O-H and O-D streching bands in the infrared have provided upper limits of $\sim 1\%$ for the solid ${\rm HDO^{\ast}/H_2O^{\ast}}$ ratio towards intermediate-mass and low-mass protostars (\citealt{Parise03}; \citealt{Dartois03}). The O-D band at 4.1 $\mu$m is weak and broad. \citet{Galvez11} estimate that the detection of HDO in amorphous ice is nearly impossible if the ${\rm HDO^{\ast}/H_2O^{\ast}}$ abundance ratio is less than a few percent.

The gas-phase ${\rm HDO/H_2O}$ ratios in hot corinos around  solar type protostars, determined recently using interferometric observations, range from $\sim 10^{-3}$ to a few percent (\citealt{Jorgensen10}; \citealt{Persson13}; \citealt{Visser13}; \citealt{Taquet13b}).  These ratios, which are likely to reflect the deuteration of water ice before desorption, conform with relatively low ${\rm HDO^{\ast}/H_2O^{\ast}}$ ratios inferred from infrared absorption measurements.

The ${\rm HDO^{\ast}/H_2O^{\ast}}$ column density ratio towards the center of our model core is $\sim 0.03$ and $\sim 0.07$ at $t=10^5$ yr and $t=10^6$ yr, respectively. These values are close to the high end of the range of gas-phase ${\rm HDO/H_2O}$ abundance ratios derived in hot corinos. In principle, the ${\rm HDO^{\ast}}$ column density in the model core becomes large enough to be determined through infrared absorption. Adopting the parameters used by \citet{Dartois03}, we estimate that the peak optical thickness of the 4.1 $\mu$m O-D absorption band from amorphous HDO ice (FWHM $\sim 0.2\,\mu$m) is about 0.02 and 0.18 at the times $t=10^5$ yr and $t=10^6$ yr, respectively.  These values are given for an offset from the centre corresponding to one half of the outer radius where the visual extinction through the core is $A_{\rm V}\sim 23$ mag. The ${\rm HDO^{\ast}/H_2O^{\ast}}$ abundance ratios at the two times quoted are $\sim 0.02$ and $\sim 0.06$. The inclusion of a low-density envelope decreases these values and increases the extinction through the cloud, but the determination of the ratio may be feasible towards a background star in a situation corresponding to a late evolutionary stage of the model core.

\section{Conclusions} \label{s:conclusions}

We have studied deuterium chemistry in starless cores with a gas-grain chemical model utilizing new chemical reaction sets for both gas phase and grain surface chemistry that include the spin states of the light species ($\rm H_2^+$, $\rm H_2$ and $\rm H_3^+$) and deuterated forms of species with up to 4 atoms. It was found that at the high densities ($n_{\rm H} \geq 10^5$\,cm$^{-3}$) and low temperatures ($\leq 10$\,K) appropriate to the centers of starless cores, HD eventually depletes from the gas phase because of surface chemistry: deuterium is efficiently locked into grain-surface HDO. Efficient HD depletion, particularly in the very centers of the cores, means that the molecular ions previously thought to be tracers of the innermost regions of starless cores, $\rm H_2D^+$ and $\rm D_2H^+$, will eventually disappear from the gas phase. It should be noted that in addition to these two ions, other deuterated species dependent on the abundance of $\rm H_2D^+$, such as $\rm DCO^+$ and $\rm N_2D^+$, show similar behavior. The timescale of HD depletion increases if chemical interaction between the gas and the dust precedes the starless core phase. In this case, part of the oxygen reservoir is locked in grain-surface $\rm H_2O$ in the beginning of the starless core phase, which limits surface $\rm HDO$ production.

While the present model predicts that HD ultimately depletes from the gas phase, our results do not imply that species such as $\rm H_2D^+$ and $\rm D_2H^+$ would be completely unusable as tracers of cold, dense gas. On the contrary, at later stages of core evolution, the $\rm H_2D^+$ and $\rm D_2H^+$ abundances increase strongly towards the core edge (i.e. lower density) and the distributions become extended, which agrees with observational evidence of extended distributions of $\rm H_2D^+$ \citep[\citealt{Vastel06}, in L1544;][in L183]{Pagani09} and $\rm D_2H^+$ \citep[in Oph H-MM1]{Parise11}. In this sense, a high observed $\rm H_2D^+$ and/or $\rm D_2H^+$ abundance toward a starless or prestellar core could (loosely) constrain the core age, although the detailed behavior of the deuteration chemistry is uncertain as it depends on various parameters, including the initial $\rm H_2(o/p)$ ratio. We note that the evaluation of the initial elemental abundances and the initial $\rm H_2(o/p)$ ratio would require detailed modeling of the core condensation phase. Also, more laboratory work on the $\rm O + H_2$ surface reaction (and on tunneling on grain surfaces in general) is required, as this reaction can significantly affect our results.


A natural extension of this work is to perform a larger-scale study of deuterium chemistry including comparisons with observations -- this will be the subject of an upcoming paper. Finally, we note that the timescale of forming grain-surface HDO, critical to the gas phase abundance of HD according to our results, depends on the surface abundances of OH and OD. As our model does not include a multilayer treatment of surface chemistry, it may overestimate the abundance of radicals available on the surface \citep[as pointed out by][]{Taquet12a}. This could have implications on HD depletion, especially if HD depletion proceeds through conversion into $\rm HDO^{\ast}$ as in our models. While the present model does not allow for it, this issue should be investigated.

\begin{acknowledgements}

We thank the anonymous referee for helpful comments that improved the paper. We also thank T.~Albertsson, S.~Cazaux, K.~Crabtree and D.~Semenov for helpful discussions. O.S. and J.H. acknowledge support from the Academy of Finland grant 132291.

\end{acknowledgements}

\appendix

\section{Details of the ortho/para separation}\label{appendixa}

In this appendix we further discuss the ortho/para separation of the OSU reaction set. As mentioned in Sect.\,\ref{sect2}, there are several reaction types where branching ratios based on spin selection rules, such as in reaction (\ref{h3+sep}), are not used; we list these special cases here. For each reaction type, we describe the assumptions behind the adopted branching ratios (if any) and give an example reaction.

\subsection{Charge transfer reactions}\label{typea1}

We assume that in charge-transferring reactions such as
\begin{equation}
{\rm H_2^+} + {\rm C_2H} \mathop{\longrightarrow}\limits^{k_{\rm A1}} {\rm C_2H^+} + {\rm H_2}
\end{equation}
the spin state of $\rm H_2^+$ (or $\rm H_3^+$) is conserved (we require that total nuclear spin in conserved in the reaction). The above reaction then separates into two reactions:
\begin{equation}
{\rm H_2^+(o)} + {\rm C_2H} \mathop{\longrightarrow}\limits^{k_{\rm A1}} {\rm C_2H^+} + {\rm H_2(o)}
\end{equation}
and
\begin{equation}
{\rm H_2^+(p)} + {\rm C_2H} \mathop{\longrightarrow}\limits^{k_{\rm A1}} {\rm C_2H^+} + {\rm H_2(p)} \, ,
\end{equation}
with the same rate coefficient $k_{\rm A1}$.

\subsection{Reactions of species other than $\rm H_2^+$ or $\rm H_3^+$ where $\rm H_2$ is created}\label{typea2}

In these reactions, neither $\rm H_2^+$ nor $\rm H_3^+$ appears as a reactant; thus, reactions such as
\begin{equation}\label{eqa1}
{\rm NH^+} + {\rm H_2O} \mathop{\longrightarrow}\limits^{k_{\rm A21}} {\rm HNO^+} + {\rm H_2}
\end{equation}
and
\begin{equation}\label{eqa2}
{\rm C_2H_3^+} + {\rm H} \mathop{\longrightarrow}\limits^{k_{\rm A22}} {\rm C_2H_2^+} + {\rm H_2}
\end{equation}
fall into this category. For these reactions, we have assumed that only para $\rm H_2$ is formed (and that, accordingly, the rate coefficient remains unchanged). Production of ortho-$\rm H_2$ requires an exothermicity of at least $\sim$170\,K (the energy difference between ortho and para states of $\rm H_2$). The exothermicity of each reaction can be calculated if the enthalpies of each species involved in the reaction are known - we have however not attempted such calculations here, and proceeded on the assumption that the required exothermicity is not reached in general.

\subsection{Reactions involving $\rm H_3^+$ where $\rm H_2$ is created}

The majority of the ion-molecule reactions of $\rm H_3^+$ are of the same type as reaction (\ref{reac1}) in the main text where $\rm H_3^+$ donates a proton and spin selection rules are applied. However, there are also reactions such as
\begin{equation}\label{eqa3}
{\rm H_3^+} + {\rm MgH} \mathop{\longrightarrow}\limits^{k_{A3}} {\rm Mg^+} + {\rm H_2} + {\rm H_2} \, ,
\end{equation}
where multiple $\rm H_2$ molecules are formed. For these reactions, we assume the following separation rules:
\begin{equation}
{\rm H_3^+(o)} + {\rm MgH} \mathop{\longrightarrow}\limits^{k_{A3}} {\rm Mg^+} + {\rm H_2(o)} + {\rm H_2(p)} \, ,
\end{equation}
\begin{equation}
{\rm H_3^+(p)} + {\rm MgH} \mathop{\longrightarrow}\limits^{{1\over2}k_{A3}} {\rm Mg^+} + {\rm H_2(o)} + {\rm H_2(p)} \, ,
\end{equation}
and
\begin{equation}
{\rm H_3^+(p)} + {\rm MgH} \mathop{\longrightarrow}\limits^{{1\over2}k_{A3}} {\rm Mg^+} + {\rm H_2(p)} + {\rm H_2(p)} \, .
\end{equation}
This logic is also applied to the (few) similar reactions that involve $\rm H_2^+$ and a molecule with several hydrogen atoms. In principle, reaction (\ref{eqa3}) can also produce $\rm H_2(o) + H_2(o)$ which can be shown by calculating the branching ratios with the method of Oka (\citeyear{Oka04}; see also \citealt{Wirstrom12}). However, branching ratios based solely on nuclear spin statistics might not be appropriate for reactions with low exothermicies, and a simplifying assumption about the branching ratios is utilized. We note that while reactions such as (\ref{eqa3}) are not among the most important reactions controlling the spin states of $\rm H_2$ and $\rm H_3^+$ in starless cores, this important issue warrants further investigation.

\subsection{Reactions involving $\rm H_3^+$ or $\rm H_2^+$ with no $\rm H_2$ produced}

These are reactions of the same type as the reaction
\begin{equation}
{\rm H_2^+} + {\rm O} \mathop{\longrightarrow}\limits^{k_{A4}} {\rm OH^+} + {\rm H} \, .
\end{equation}
In these cases, ortho and para states are destroyed with equal rates, so that the above reaction separates into
\begin{equation}
{\rm H_2^+(o)} + {\rm O} \mathop{\longrightarrow}\limits^{k_{A4}} {\rm OH^+} + {\rm H} \, ,
\end{equation}
and
\begin{equation}
{\rm H_2^+(p)} + {\rm O} \mathop{\longrightarrow}\limits^{k_{A4}} {\rm OH^+} + {\rm H} \, .
\end{equation}

\subsection{Other reactions}

In addition to the reaction types presented above, there are also a number of reactions that do not fall into the general categories discussed above. Whenever this is the case, we apply a custom logic unique to each reaction that is, as far as possible, consistent with the above general cases. For example, the reaction
\begin{equation}
{\rm H_2^+} + {\rm H_2S} \mathop{\longrightarrow}\limits^{k_{A5}} {\rm HS^+} + {\rm H} + {\rm H_2}
\end{equation}
separates into the reactions
\begin{equation}
{\rm H_2^+(o)} + {\rm H_2S} \mathop{\longrightarrow}\limits {\rm HS^+} + {\rm H} + {\rm H_2(o/p)} \, ,
\end{equation}
and
\begin{equation}
{\rm H_2^+(p)} + {\rm H_2S} \mathop{\longrightarrow}\limits {\rm HS^+} + {\rm H} + {\rm H_2(o/p)} \, ,
\end{equation}
with 1:1 branching ratios for both reactions. For the former reaction, the method of \citet{Oka04} yields an $\rm H_2$ ortho/para ratio of 2:1 (and 1:1 for the latter reaction), when $\rm H_2S$ can exist in either ortho or para form. But since we do not know the spin state of $\rm H_2S$, we assume that it exists totally in para form and make the above simplifying assumption.

\bibliographystyle{aa}
\bibliography{20922.bib}

\end{document}